\documentclass[runningheads]{svmult}

\usepackage{makeidx}   
\usepackage{graphicx}  
\usepackage{subeqnar}  
\usepackage{multicol}  
\usepackage{physprbb}  
\makeindex             

\begin{document}
\title*{Observations of Gamma-ray Bursts\protect\newline 
with the Rossi X-ray Timing Explorer} 
\toctitle{Observations of Gamma-ray Bursts
\protect\newline with the Rossi X-ray Timing Explorer}
\titlerunning{Observations of GRB with RXTE}

\author{Hale Bradt\inst{1}
\and Alan M. Levine\inst{1}
\and Francis E. Marshall\inst{2}
\and Ronald A. Remillard\inst{1}
\and Donald A. Smith\inst{3}
\and Toshi Takeshima\inst{2}$^{,}$\inst{4}}

\authorrunning{Bradt et al.}

\institute{Massachusetts Institute of Technology, Cambridge MA 02139-4307, USA, 
Room 37-587; bradt@mit.edu
\and Code 662, Goddard Space Flight Center, NASA, Greenbelt MD 20771, USA
\and 2477 Randall Laboratory, University of Michigan, Ann Arbor MI 48109, USA
\and Present address: NASDA, Tokyo, Japan 105-8060}

\maketitle

\begin{abstract}

The role of the Rossi X-ray Timing Explorer (RXTE) in the study of 
Gamma-ray Bursts (GRBs) is reviewed. Through April 2001, the All-Sky Monitor 
(ASM) 
and the Proportional Counter Array (PCA) instruments have detected 30 GRBs. In  16 cases, an early celestial position was released to the community, sometimes in conjunction with IPN results. The subsequent optical and radio 
searches led to the detection of 5 x-ray afterglows, to at least 6 optical or radio 
afterglows, to 3 of the 17 secure redshifts known at this writing, and to 2 other likely 
redshifts. The decay curves of early x-ray afterglows have been measured. The rapid 
determination of the location of GRB 970828 and the absence of optical afterglow at 
that 
position gave one of the first indications that GRBs occur in star-forming 
regions~\cite{gro98}. The location of GRB 000301C led to the determination of a 
break in the 
optical decay rate~\cite{rho01} which is evidence for a jet, and to variability in the 
optical 
light curve that could represent gravitational lensing~\cite{gar00}. X-ray light curves 
of GRB 
from the ASM in conjunction with gamma-ray light curves exhibit striking differences 
in 
different bands and may reveal the commencement of the x-ray 
afterglow~\cite{smi01}. 
\end{abstract}

\section{Introduction}
The Rossi X-ray Timing Explorer (RXTE)~\cite{bra93} has played a highly significant 
role in the explosive growth of GRB studies in the afterglow era. The 
RXTE contributions have been possible because of its ability to rapidly point to a new 
target, the sensitivity and high-energy response (2 $-$ 60 keV) of the 
PCA~\cite{jah96}, and the wide-field position-determining capability 
of the ASM~\cite{lev96}. This overview encompasses work through April 2001.

\section{Observations of afterglows with the PCA}

\begin{figure} [t]
\begin{center}
\includegraphics[width=1.0\textwidth] {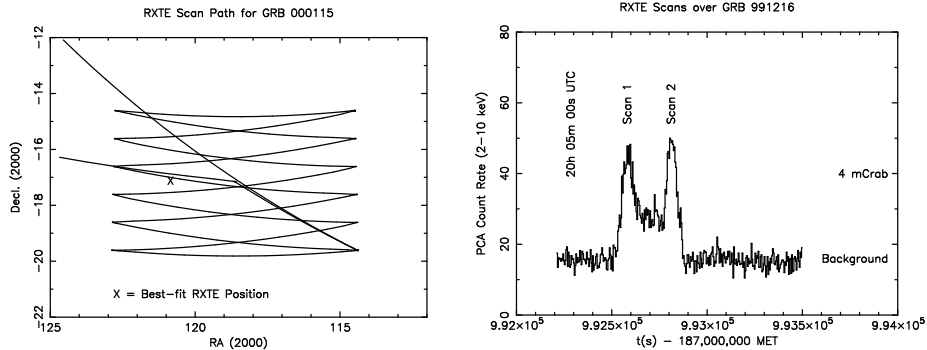}
\end{center}
\caption{Typical raster scans by RXTE PCA. Left: the scan track used for 
GRB 000115. Right: PCA counting rate vs. time for the raster scan of GRB 991216}
\label{rast}
\end{figure}

Searches for early x-ray afterglows with the PCA have been accomplished mostly by 
means of rapid pointings (few hours) of RXTE that direct the PCA FOV toward the 
near real 
time BATSE ``Locburst'' position, when the position has uncertainty 
less than a few degrees. In addition early GRB positions from the ASM, BeppoSAX, 
and the Interplanetary Network (IPN) have been used as targets. Raster scans of the 
region with the 1.0 deg (FWHM) PCA field of view are used to carry out the search; a 
detection usually can be localized to a few arc minutes; Fig.~\ref{rast}. Such 
searches by the RXTE/PCA are dependent on the schedule of upcoming command 
contacts, 
and upon the burst location relative to the current celestial pointing direction of RXTE. 
The 
PCA typically reached its target within a few hours after notification of the burst 
location.

\begin{table}
\caption{PCA detections of x-ray afterglows}
\begin{center}
\renewcommand{\arraystretch}{1.4}
\setlength\tabcolsep{5pt}
\begin{tabular}{llllp{4.0cm}}
\hline\noalign{\smallskip}

GRB & BATSE peak flux & PCA flux & X-ray & Comments$^{\mathrm b}$\\
& cm$^{-2}$ s$^{-1}$ & @ 3 h & decay &\\
& (50 $-$ 300 keV) & (mCrab) & index$^{\mathrm a}$ &\\

\noalign{\smallskip}
\hline
\noalign{\smallskip}

970616* & & 0.5 & & $3\sigma$ PCA detection\\
970828* & & 1.0 & 1.6 & See Table~\ref{table2}\\
990506* & 19 & 3 & 0.9 & Radio; no opt afterglow; host at $z = 1.3$\\
991216* & 68 & 8 & 1.8 & Opt. and radio afterglow; jet in ISM, $z > 
1.02$?\\
000115* & 57 & 2 & 1.9 & No opt. afterglow\\
\noalign{\smallskip}
\hline
\noalign{\smallskip}
\end{tabular}
\end{center}

$^{\mathrm a}$~Spectral indices are from PCA data alone. \\
$^{\mathrm b}$~See references elsewhere in this report and also on J. Greiner web 
page: http://www.aip.de/$\sim$jcg/grbgen.html \\
{*}~Prompt position notice provided to community
\label{table1}
\end{table}

\begin{figure} [t!]
\begin{center}
\includegraphics[width=1.0\textwidth] {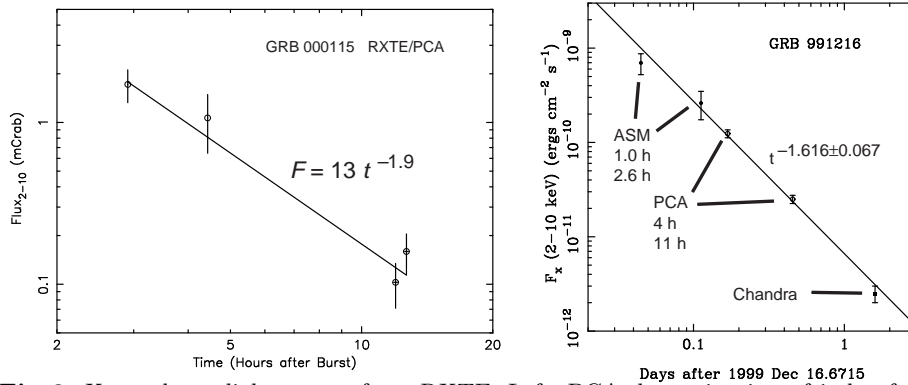}
\end{center}
\caption{X-ray decay light curves from RXTE. Left: PCA determination of index for 
GRB 000115. Right: ASM/PCA/Chandra decay of GRB 991216, reported in Halpern et 
al.~\cite{hal00}}
\label{decay}
\end{figure}

There have been 31 attempts to view x-ray afterglows with the PCA. Of these 24 
were 
triggered by BATSE Locburst notices and 7 by one or more of the RXTE/ASM, 
BeppoSAX, and the IPN. Five x-ray afterglows were detected, 
four derived from Locburst and one from the ASM; Table~\ref{table1}. In each case 
celestial
positions were promptly reported to the community. 
Highlights of PCA detections and follow-on measurements are:
\begin{itemize}
\item
Radio afterglow in GRB 990506, but with no optical~\cite{tay00}; host galaxy has 
$z = 1.3$~\cite{blo01}.
\item
Optical afterglow in GRB 991216 with a break in the decay curve indicating a jet in 
interstellar 
medium~\cite{hal00}, radio emission~\cite{fra00}, and a reported lower-limit to the 
redshift, $z > 1.02$~\cite{vre99}.
\item
X-ray decay indices of 4 bursts; see Table~\ref{table1} and Fig.~\ref{decay}.
\end{itemize}

\section{Observation of GRBs with the ASM}
The ASM, with an instantaneous field 
of view of $\sim3000$ sq. deg. and $\sim$40\% duty cycle, has upon occasion serendipitously observed a burst in 
the 
FOV of one of 
its three Scanning Shadow Cameras (SSCs). A detection in a single camera yields a 
line of position of a few arc minutes by a few degrees. Such an error region can be 
reduced with IPN lines of position or with PCA scans. Less frequently, a GRB is 
detected in  
both of the two azimuthal cameras that have overlapping FOVs ($\sim$300 sq. deg.). 
This 
yields crossed lines of position and an error region of order 10 sq arc min. This is 
sufficient for 
radio and optical searches, though PCA or IPN results can further refine 
the position. 

The rate of GRB detection in the overlapping FOVs was initially expected to be 
$\sim$1 per  
year. The actual rate (see ``2-SSC'' events in Tables~\ref{table2} and \ref{table3}) 
turns 
out to be about twice this because (1) the x-ray portion of the bursts tends to be 
longer than 
the gamma-ray portion and (2) the ASM rotates 
every 100 s which shifts the FOVs 6$^\circ$ on the sky. This shift can bring a FOV 
onto a long-duration 
burst. It can also bring a camera onto a burst that had just 
been observed with the other azimuthal camera, thus providing crossed lines of 
position. 

In 5.2 years of operation, up to the present (April 2001) a total of 27 probable GRBs 
have 
been detected in the ASM. Of these 24 are confirmed as GRBs by gamma-ray 
detectors on 
other satellites; Table~\ref{table2}. Ten of these were detected in 2 SSCs and three 
were located in 
a recent global reprocessing of the entire data base. The ASM also recorded at least 
three 
other events for which there has been no reported gamma burst from other satellites; 
Table~\ref{table3}. Each of the listed events has a hard x-ray spectrum, is at  
high galactic latitude, and is a one-time detection.  Thus they could be x-ray 
bright GRBs. Two of these appear to be afterglow detections as no fast variability is 
apparent during the 100-s ASM exposures. 

The ASM has been providing rapid celestial positions to the community since the 
commencement of the afterglow era in 1997; to this date 12 such positions have been 
reported, sometimes in conjunction with IPN reports. Of these, four were observed in 
two 
cameras and hence yielded small uncertainties in two dimensions. 
\begin{table}[t!]
\caption{ASM detections of confirmed GRBs}
\begin{center}
\renewcommand{\arraystretch}{1.4}
\setlength\tabcolsep{5pt}
\begin{tabular}{llp{9.0cm}}
\hline\noalign{\smallskip}

GRB$^{\mathrm a}$  & Confirming & Comments$^{\mathrm c}$$^{,}$$^{\mathrm d}$\\
& satellite$^{\mathrm b}$ &\\

\noalign{\smallskip}
\hline
\noalign{\smallskip}

960228  &  k &   2 SSCs; found in reprocessing\\
960416 &  b k u &   2 SSCs; soft intermediate x-ray peak; extended x-ray tail\\
960524 & b & 2 SSCs; afterglow in ASM 25 min. after burst, fnd. in reproc.\\
960529 &  k &   2 SSCs\\
960610 &  b &   2 SSCs; found in reproc.\\
960727 &  k u &   1 peak; extended x-ray tail\\
961002 & k u &   1 peak; extended x-ray tail\\
961019 &  b k u &   Delayed x-ray peak; x-ray tail\\
961029 &  k &   Limited x-ray data\\
961216 &  b k &   Poor ASM position\\
961230 &  u &   2 SSCs; weak x ray\\
970815* &  b k s u &  2 SSCs; strong tertiary x-ray peak\\
970828* &  b k o u &   2 SSCs; no opt. afterglow; z = 0.958?, 0.33? (x-ray line); rad.?\\
971024* &  b k  &   Weak; position uncertain\\
971214* &  b k n s u &  Single longer x-ray peak; OT; z = 3.42 from BSAX position\\
980703* &  b k u &   2 SSCs; OT; z = 0.966; long $\gamma$/x tails; radio\\
981220* &  k s u &   8 Crab (5$-$12 keV); $\gamma$ peaks smeared in x~rays\\
990308* &  b k u  &  OT; host galaxy underluminous or distant\\
991216 & b n u   &   OT; x-ray afterglow at opt. pos. at 1.0 h, 2.6 h (ASM)\\
000301C* &  k n u &   3 Crab (5$-$12 keV); OT; z = 2.04; rapid opt. var. 
(lensing?)\\
000508* &  b u & 2 Crab; emerged from earth occultation\\ 
001025* &  n u &   $\sim$4 Crab (5$-$12 keV); no optical \\
010126* &  k n u &   $\sim$5 Crab (5$-$12 keV); no optical\\
010324* & s u  &   2 SSCs; det. 360 s after GRB onset; no det. rapid var. in ASM, but 
decays 
factor of 4 in 100 s (afterglow?); no opt.\\

\noalign{\smallskip}
\hline
\noalign{\smallskip}
\end{tabular}
\end{center}

$^{\mathrm a}$~The GRBs detected in 1996 were located in archival searches.\\ 
$^{\mathrm b}$~b=BATSE; k=KONUS; n=NEAR;  o=SROSS-C; s=BSAX; 
u=Ulysses\\
$^{\mathrm c}$~See refs. herein and on J. Greiner: http://www.aip.de/$\sim$jcg/grbgen.html \\
$^{\mathrm d}$~OT $=$ optical transient\\
{*} Prompt position notice provided to community
\label{table2}
\end{table}

The nearly continuous telemetry stream from RXTE, via the TDRSS 
satellites, makes possible rapid position determinations, in principle less than an hour 
and in 
practice usually within a few hours of the burst occurrence time in the RXTE detectors. 
The events were typically confirmed as due to GRBs in near real time by a BATSE 
event 
and thus could be released rapidly with high confidence, well before IPN positions 
became 
available. In one case, GRB 970828, the ASM position determination made possible 
acquisition of the burst by the PCA only 3.6 h after the burst occurrence at RXTE.

The ASM positions have led to 5 optical afterglows and 2 to 4 redshifts. The range in 
the 
latter number depends upon the confidence in the determined redshift and the role the 
ASM played in instigating the search. As noted, the one-camera detections required 
other information (IPN, SAX, or PCA) to permit fruitful optical and radio searches. In 
the case of GRB 991216, the ASM did not catch the burst itself, 
but did capture the early afterglow at 1.0 h and 2.6 h after the burst. The PCA 
measured the afterglow of this same burst at 4 h and 11 h. Together with a Chandra 
observation, these provided an excellent measure of the very early x-ray decay of a 
GRB~\cite{hal00}; Fig.~\ref{decay}. 
Notable accomplishments that stemmed from ASM results are:

\begin{itemize}
\item
No optical afterglow was detected in GRB 970828 possibly due to its occurring in a 
star forming region~\cite{gro98}. Variability in the x-ray afterglow decay was found 
with ASCA~\cite{yos00}.
\item
Redshifts were obtained for the host galaxy of GRB 980703 ($z = 
0.966$)~\cite{djo98} and for the afterglow of GRB 000301C ($z = 2.04$)~\cite{jen00} 
\item
The optical afterglow decay of GRB 000301C showed a clear break~\cite{rho01} and 
remarkable rapid optical variability on time scales of $\sim$1/2 
day~\cite{sag00}~\cite{mas00} suggestive of a microlensing event~\cite{gar00}.
\item
The x-ray afterglows of GRB 960524 and GRB 991216 were measured by the ASM 
only 25 min and 1.0 h after the burst, respectively (see above). 
\item
Coincident x-ray (ASM) and gamma-ray light curves of 15 GRBs show marked 
differences between low and high energies; Fig.~\ref{lc}.
\end{itemize}


\begin{table}
\caption{ASM GRB not confirmed by gamma detectors}
\begin{center}
\renewcommand{\arraystretch}{1.4}
\setlength\tabcolsep{5pt}
\begin{tabular}{llp{2.0 cm}llp{3.0cm}}
\hline\noalign{\smallskip}

Date & $\alpha, \delta^{\mathrm a}$ & Start (MJD)& $I_{max}$ & Gal. & 
Comments\\
& (J2000) & (interval) & (mC) & lat. & \\

\noalign{\smallskip}
\hline
\noalign{\smallskip}

961225 & 154.97, 64.04 & 50442.2620 ($>$90s) & 360 & +46$^\circ$ & 2 SSCs; flat 
l.c.; afterglow?\\

970123 & 184.68, $-$21.20 & 50471.0707 ($>$190 s) & 62 & +41$^\circ$ & 2 SSCs; 
ramping down; afterglow?\\

000913 & 16.355, $-$16.008 & 51800.6711 & 700 & $-78^\circ$ & \\  

\noalign{\smallskip}
\hline
\noalign{\smallskip}
\end{tabular}
\end{center}
$^{\mathrm a}$ $\sim$90\% errors: 0.2$^\circ$ for the first two bursts. For the third: 
line length = 2.50$^\circ$, line width = 0.090$^\circ$, pos. angle = $-$36.930$^\circ$
\label{table3}
\end{table}

\section{Coincident x/gamma light curves}

The ASM provides light curves for GRB detections in 
three x-ray energy bands: 1.5 $-$ 3 keV, 3 $-$ 5 keV, and 5 $-$ 12 keV.  In addition, 
the ASM 
time series data mode may reveal temporal variability in each channel down to a time 
resolution of 0.125 s if the GRB is the dominant x-ray source in a camera's FOV. 
These 
data provide valuable information in a band not typically monitored by gamma-ray 
experiments. (But note, e.g., the x-ray detections of GRBs ($<10$ keV) with
Ginga~\cite{oga91} and with the currently operational WFC on 
BeppoSAX~\cite{fro00}.)

\begin{figure} [thb]
\begin{center}
\includegraphics[width=.5\textwidth] {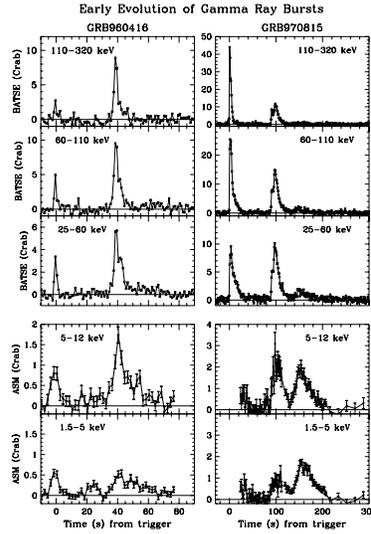}
\end{center}
\caption{Light curves in 3 BATSE energy bands and 2 ASM bands of GRB 960416  
and GRB 970815, from Smith et al.~\cite{smi01}}
\label{lc}
\end{figure}

The ASM in conjunction with gamma-ray detectors on BATSE, BeppoSAX, Ulysses, 
Konus, and NEAR have yielded multifrequency light curves over the range 1.5 to 
$>$300 keV. Fifteen of these have been collected and analyzed by Smith et 
al.~\cite{smi01}. They show diverse morphologies with striking differences between the x-ray and gamma-ray bands; Fig.~\ref{lc}. For example, the pronounced 3rd ASM peak of GRB 970815 is not detected in the upper BATSE energy bands. This peak could represent the beginning of the afterglow in the external shock model~\cite{mes97}. This supposition is supported by (1) the $\sim8-$s delay of the peak time in the highest ASM energy channel ($\sim$7 keV) relative to that of the lowest channel ($\sim$2.25 keV) and (2) the achromatic decay of this peak, with decay indices in each of the 3 ASM energy channels consistent with $\alpha = 1.3 \pm 0.1$. 

\begin{figure} [thb]
\begin{center}
\includegraphics[width =1.0\textwidth] {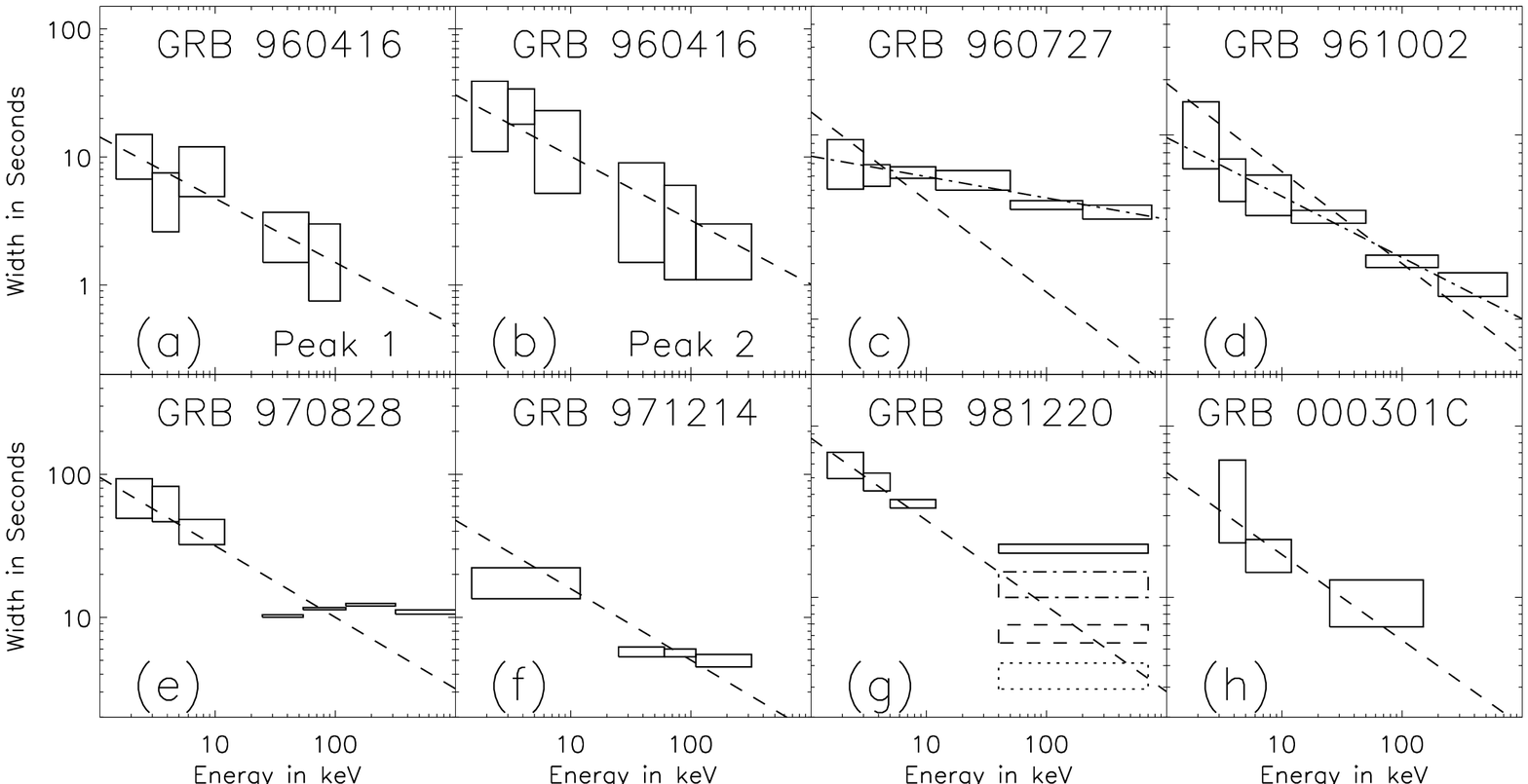}
\end{center}
\caption{Peak width vs. energy for seven GRBs. The dashed lines indicate a 
logarithmic slope of $-$0.5, from Smith et al.~\cite{smi01}}
\label{width}
\end{figure}

Smith et al.~\cite{smi01} also studied the durations of gamma bursts as a function of energy and compared them with the $E^{-0.5}$ 
power law expected from an origin in synchrotron radiation~\cite{pir99}. Examples that match well are the two distinct peaks of GRB 960416 taken separately (Figs.~\ref{lc},~\ref{width}a,b) and the single peak (see \cite{smi01}) of GRB 000301C (Fig.~\ref{width}h). The more complex bursts (see light curves in \cite{smi01}) are not consistent with this prediction. Some have flatter curves which widen more slowly than expected with decreasing energy; Fig.~\ref{width}c,d. Other complex bursts show a flat curve at high energies but an excess at low energies; Fig.~\ref{width}e,f. These inconsistencies are probably indicative of complex interactions that violate the assumption of a single infusion of energy followed by cooling through radiation.

\section{RXTE burst studies in the HETE era}
The RXTE will continue to contribute occasional additional positions, light curves, and 
decay indices of GRBs with increased reliance on ASM self-triggered events and on IPN crossing lines of positions and confirmations, now that BATSE is no longer in orbit. Most important, the RXTE/PCA has the unique potential to acquire a HETE burst extremely rapidly, say within  1/2 hour of the burst and to study the temporal activity in the afterglow with high statistics, a domain not heretofore explored. 

\section*{Acknowledgments}
We acknowledge the contributions of the many individuals on the ASM and PCA 
teams, the observers whose data are included in Fig.~\ref{lc}, and NASA for the 
continued operation of RXTE.

\end{document}